\documentclass{llncs}

\usepackage{graphics,epsfig}
\input{psfig.sty}

\begin{document}

\pagestyle{empty}

\mainmatter

\title{Contextual Random Boolean Networks}

\titlerunning{Contextual Random Boolean Networks}

\author{Carlos Gershenson
\and Jan Broekaert \and
Diederik Aerts}

\authorrunning{Carlos Gershenson et al.}

\institute{Centrum Leo Apostel, Vrije Universiteit Brussel,
Krijgskundestraat 33,\\
Brussels, 1160, Belgium\\
\email{\{cgershen, jbroekae, diraerts,\}@vub.ac.be}\\
\texttt{http://www.vub.ac.be/CLEA}}

\maketitle

\begin{abstract}
We propose the use of Deterministic Generalized Asynchro\-nous Random Boolean Networks \cite{gershenson01} as
models of contextual deterministic discrete dynamical systems. We show that changes in the context have
drastic effects on the global properties of the same networks, namely the average number of attractors and the
average percentage of states in attractors. We introduce the situation where we lack knowledge on the context
as a more realistic model for contextual dynamical systems. We notice that this makes the network
non-deterministic in a specific way, namely introducing a non-Kolmogorovian quantum-like structure for the
modelling of the network \cite{aerts01}. In this case, for example, a state of the network has the
potentiality (probability) of collapsing into different attractors, depending on the specific form of lack of
knowledge on the context.
\end{abstract}

\section{Introduction}

Random Boolean Networks (RBNs) have been used to model a variety of phenomena and have been
widely studied \cite{wuensche01,wuensche02,aldana01}. Especially, they were used by Stuart Kauffman
\cite{kauffman01} to study genetic regulatory networks and to correlate the number of human cell types to the
attractors of a RBN with a number of nodes similar to the number of genes in the DNA. This is interesting,
because one does not assume any function in the RBN, these are generated randomly. However, this approach was
criticized because it assumed that the dynamics of the DNA would be discrete and synchronous in time
\cite{harvey01,dipaolo01}. The proposed alternative, non-deterministic asynchronous updating, did not give
encouraging results, since the networks change drastically their properties due to the non-determinism. We proposed
another type of updating: asynchronous but deterministic \cite{gershenson01}. This is achieved by setting parameters
to determine the period of the updating of each node. The problem now lies in how to find or study different
updating periods.

Our approach in the present paper consists in the following: we see the parameters of the updating periods of the
nodes as a \emph{context} of the network. We then study what happens when we change this context for a given
network, \emph{i.e.} allow smaller or greater periods. We find out that the properties of the network change
drastically if we change the context.

We mentioned already that the non-deterministic asynchronous updating led to effects of changes that are too drastic
due to the non structured nature of the non-determinism. In our approach we want to introduce non-determinism, but
in a very specific and structured way: namely non-determinism related to a Òlack of knowledgeÓ on the specific
context (updating period) that we consider. The idea is that we start by introducing contexts that influence the
network deterministically, and call these contexts \emph{pure contexts}. Under the effect of such a pure context the
network still behaves completely deterministically, but each pure context will generate a different behaviour for
the network. These pure contexts are very idealized models for context in the real world. If we want to model the
effect of a real context, we have to take into account that we generally lack knowledge on how this real context
interacts with the system, because of the presence of random and unpredictable fluctuations. Hence we model such a
real context by introducing the idea of \emph{mixed context}, where a mixed context is a statistical mixture of a
set of pure contexts. Concretely this means that we describe a mixed context by a probability measure on the set of
pure contexts. We will show that the effect of this structured type of non-determinism is that states of the network
become \emph{potentiality states}, in the sense that instead of a state evolving deterministically to one of the
attractors, such a state will now be attributed a certain probability (potentiality) to \emph{collapse} to one of
the different attractors under the influence of a certain context, the probabilities being determined by the weights
of the different pure contexts within the considered mixed context. 

So the behaviour of the network becomes
non-deterministic when we introduce mixed contexts, but it is much more structured than in asynchronous RBNs where
anything can happen, and therefore it can be well studied. The structure of the probability model that arises by
means of the introduction of mixed contexts has been studied in detail, and it can be proven that this structure is
non-Kolmogorovian \cite{aerts01}. 

We want to study contextual Random Boolean Networks by introducing the effect of
the context in the way we just explained because it is a realistic model for the effect of real contexts. There is
however a second reason, namely, this type of non-deterministic contextual influence has been studied in great
detail in our Brussels research group, leading to a modelling of contextuality that is quantum-mechanic-like. What
we mean more specifically is that the mathematical structure that emerges from a situation where non-determinism is
introduced as a consequence of the presence of mixed contexts is a quantum mechanical mathematical
structure \cite{aerts01,aerts02,aerts03,aerts04,aerts05}. Furthermore this type of contextual models have been used
in different ways, mostly to model the contextuality in situations of cognition
\cite{aertsaerts01,aertsetal01,aertsetal02,aertsetal03,gaboraaerts01}. There are reasons to believe that this type
of contextuality is also present in biological systems \cite{gabora01,aertsetal04}, and that is the reason that we
introduce it for Random Boolean Networks.

In the following section, we make a brief review of different types of RBNs, according to their updating
scheme \cite{gershenson01}. In Section 3, we analyse the effects of the change of pure context in the statistical
properties of a RBN. In Section 4, we use a concrete example of a small contextual RBN to note the
properties of a mixed context. We draw conclusions and future lines of research in Section 5.

\section{Background}

A Random Boolean Network \cite{kauffman02,kauffman01} can be seen as a generalization of a boolean cellular automata. It
also consists of $n$ nodes, each with $k$ connecting inputs, only that the connections are not restricted to neighbours,
but they can be to any other node in the network. The topology is generated randomly, but it remains fixed during
the dynamics of the network. These dynamics are determined by boolean rules also generated randomly in lookup
tables. The state of a node will depend on the states of the nodes to which it is connected, \emph{i.e.} its
inputs.

It is interesting to study the general properties of RBNs with different number of nodes and connections, making
statistical analyses, because we can obtain general properties of the dynamics of a family of networks, without
assuming the rules of the dynamics. Nevertheless, it has been shown that these properties can change drastically
depending on how we update the nodes \cite{harvey01,gershenson01,cornforthetal01}.

We have proposed a classification of RBNs \cite{gershenson01} according to the different updating schemes that a network
might have:

\textbf{Classical Random Boolean Networks} (CRBNs) \cite{kauffman02,kauffman01}. The updating is synchronous and
deterministic: each node takes its value at time t+1 from the values of the nodes connected to it at time t.

\textbf{Asynchronous Random Boolean Networks} (ARBNs) \cite{harvey01}. The updating is asynchronous, but also
non-deterministic. Each time step only one node is picked at random and updated.

\textbf{Deterministic Asynchronous Random Boolean Networks} (DARBNs) \cite{gershenson01}. The updating is asynchronous and
deterministic. To achieve this, we introduce two parameters per node: $p$ and $q$, so that the node will be updated when
time modulus $p$ equals $q$. Therefore, $p$ can be seen as the period and $q$ as the translation in time of the node update.
If more than one node is updated at one time step, the network is actualized in the same order after each node
update.

\textbf{Generalized Asynchronous Random Boolean Networks} (GARBNs) \cite{gershenson01}. The updating is non-deterministic,
but semi-synchronous. We select randomly at each time step some nodes to update, and these are updated synchronously.

\textbf{Deterministic Generalized Asynchronous Random Boolean Networks} (DGARBNs) \cite{gershenson01}. The updating is
deterministic and semi-synchronous. We also use parameters $p$ and $q$ in each node to determine the period and
translation of each update, but the nodes which should be updated at time $t$ do so synchronously.

In the deterministic cases (CRBN, DARBN, DGARBN), there can be cyclic attractors (when the dynamic of the network is
cycled in a subset of the state space) and point attractors (when a single state ``traps" the dynamics). On the other
hand, for the non-deterministic cases (ARBN, GARBN) there are point attractors, and loose attractors (a subset of
the state space drags the dynamic, but this can follow several patterns inside this subset, since we do not know
which node will be updated at a given time). To our knowledge loose attractors have not been studied, since it is
not trivial to find them.\\
We will use DGARBNs to study the change of context in RBNs.\\

\noindent {\bf Resources:} We have developed a software laboratory, ÒRBNLabÓ, for studying the properties of different types of RBNs. It
is available to the public, for use (via browser) and download (Java source code included) at
\\ http://student.vub.ac.be/cgershen/rbn. The results presented in this paper were obtained using RBNLab.

\section{Changing Contexts in DGARBNs}
We consider the set of all $p$'s and $q$'s as the \emph{context} of the DGARBN. This is because in real networks
some external factors, such as temperature or tension \cite{ingber01}, can produce a change in the updating regularity of
the nodes. The external factors are thus reflected in the updating periods, which are represented with the sets of
parameters of $p$'s and $q$'s. We should note that this is a temporal context, but might reflect other types of
contexts that affect a system.

After generating randomly the topology and rules of a network of given $n$ and $k$, we randomly generate contexts
according to a parameter maxP, which indicates the maximum allowed period for a node. The $p$'s are random integers
between 1 and maxP, while $q$'s are also random integers, but between 0 and maxP-1. We can see that the case maxP=1
gives us CRBNs (fully synchronous).

It could be argued that it is equivalent to study CRBNs that would consider the state and context of a DGARBN as a
state of a more complex CRBN, since we have shown that any deterministic asynchronous RBN can be mapped into a CRBN
of higher complexity \cite{gershenson01}. This is achieved by adding nodes (encoding the context) connected to every other
node in the network. Remark that a network of given $n$, $k$, and maxP can be converted into a CRBN of, $\forall i$,
$n+{\rm ceiling}(\log 2({\rm LCM}( p_i)))$ and  $k + {\rm ceiling}(\log2({\rm LCM}( p_i)))$. First, it is conceptually
more clear to divide the network and its context, because changing the context in a DGARBN or DARBN is easy and not
so in its CRBN correlate. Moreover, the family of CRBNs of extended $n$ and $k$ is much larger than the one of
contextual RBNs, so we speculate  they have different statistical properties.

About statistical properties, we should note that all types of RBNs tend to have different properties in theory
(exact solutions) than in practice (statistics). This was shown already by Harvey and Bossomaier \cite{harvey01} for ARBNs.
For these type of RBN, the exact solution (expected if one would exhaust all the RBN family, but this is unfeasible
computationally) for the average number of attractors is exactly one, for any type of topology (combinations of $n$
and $k$). However, the statistics give us a different result. In some networks ($k=3$) the distribution of attractors
is such that these tend to ÒhideÓ (few networks with several attractors, many with none). We are aware that in all
RBNs these divergences between theory and practice are common. But we are interested more in statistical
properties than in exact solutions, because we aim at finding a network of
certain properties, without exhausting the network space.

\bigskip
\noindent
{\bf Method:}
We generated one thousand DGARBNs for each given $n$ and $k$, and explored the properties of the same networks as we varied maxP. 
Remark that for $n>10$, for computational resources issues, we sampled only 100 networks. Results for $n>10$ should thus be considered illustrative.
To find the properties of the networks, we explore all possible initial conditions, running the network for ten thousand steps expecting that the
network will reach an attractor. The attractors take into account the state and the context, so for example a
point attractor (period one in CRBN) will be considered as having a period equal to the least common multiple
of all $p$'s. This is to explore all the possible combinations of updates in the nodes.

\bigskip
\noindent
{\bf Results:}
The average number of attractors of the explored networks of different $n$ and $k$ values can be
seen in Figure 1.  Figure 2 shows the results for different $n$'s and $k=3$.
\begin{figure}
\centerline{\psfig{figure=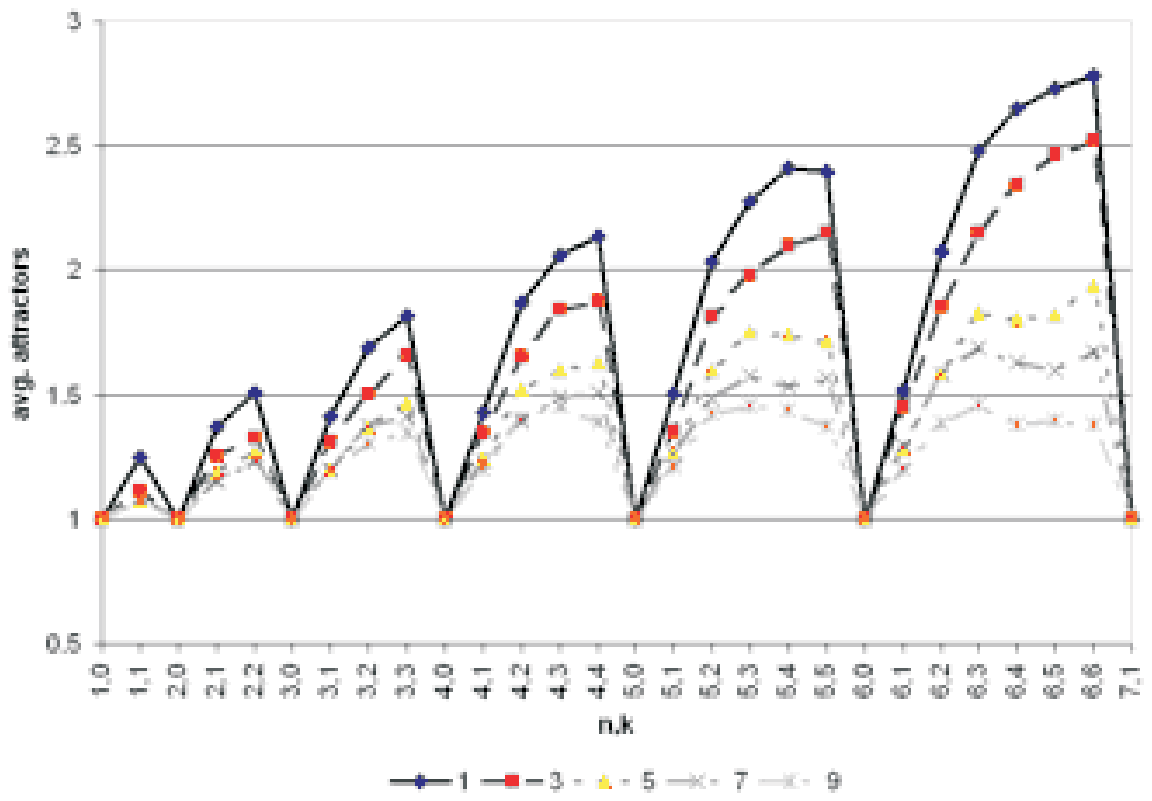,height=4.5cm}}
\caption{Average number of attractors varying maxP}
\end{figure}

\begin{figure}
\centerline{\psfig{figure=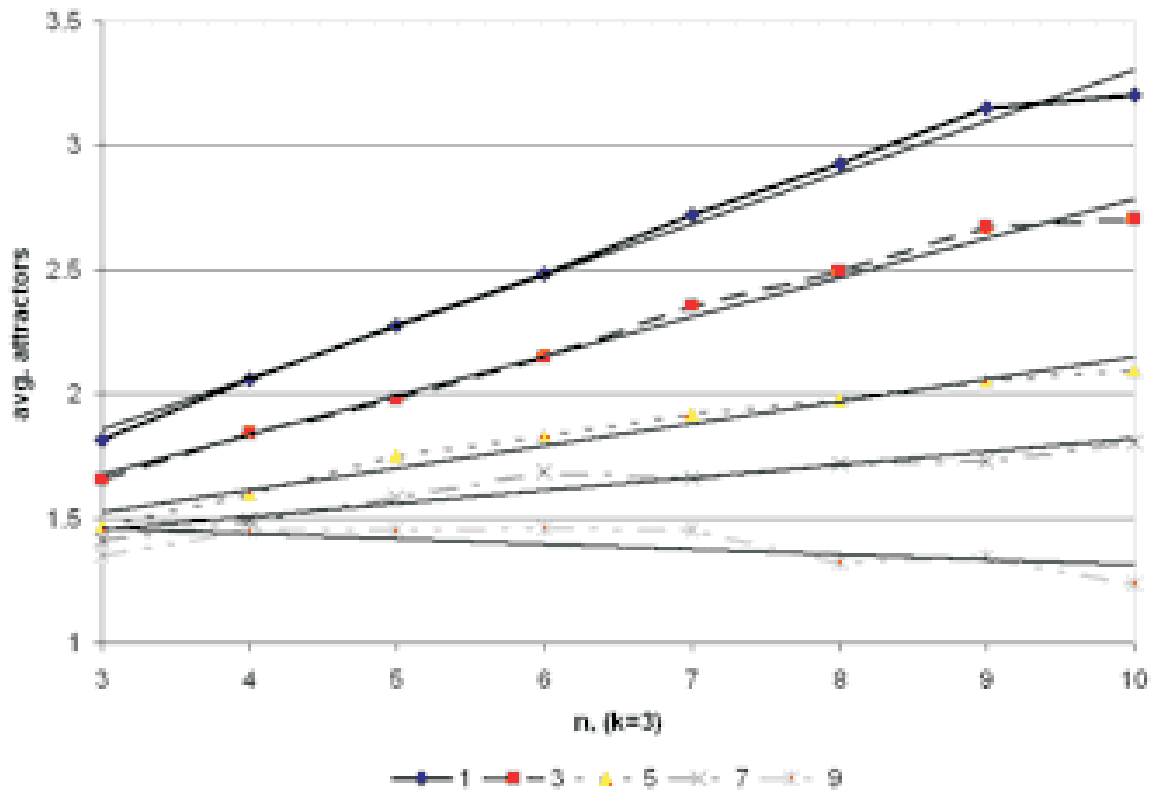,height=4.5cm}}
\caption{Average number of attractors varying maxP}
\end{figure}
The first thing we can see by observing these graphs is that the context change has drastic consequences on the
average number of attractors. Moreover, comparing these results with the ones of Gershenson \cite{gershenson01} contexts
of small maxP clearly yield network properties very similar to synchronous RBNs (CRBNs). Still, as we increase maxP,
the properties begin to look more like the ones of asynchronous non-deterministic RBNs (ARBNs and GARBNs).
Another thing we can observe is that even when the average number of attractors for low maxP has a linear increment
proportional to n, this decreases and becomes even decrement for high maxP. This means that for large networks, the
differences in the number of attractors will grow considerably as we change the context. The attractors of the same
networks collapse as we change the context and allow larger periods (increase maxP). The inverse results are obtained in CRBNs including context as part of their state, since the average number of attractors in CRBNs is increased with $n$ and $k$. We also know from Gershenson \cite{gershenson01} that the point attractors are the same for every type of RBN, so the attractors that collapse are not these ones, but the cycle
attractors.
\begin{figure}
\centerline{\psfig{figure=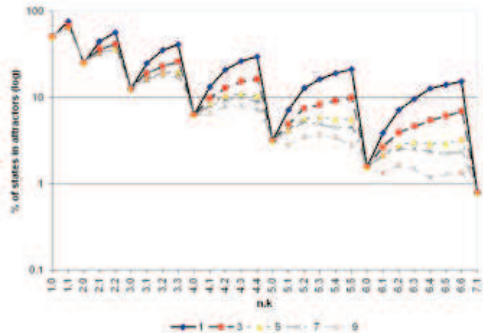,height=4.5cm}}
\caption{Percentage of states in attractors varying maxP (log scale)}
\end{figure}
The percentage of states of the networks that are part of an attractor in logarithmic scale can be seen in Figure 3
for different $n$'s and $k$'s and for $k=3$ in Figure 4.

We can see that all the percentages of states in attractors decrease exponentially as we increase $n$. But the
percentage of states in attractors decreases slower for low maxP than for a high one. This also indicates that the
relative differences among contexts will be greater for larger networks.
\begin{figure}
\centerline{\psfig{figure=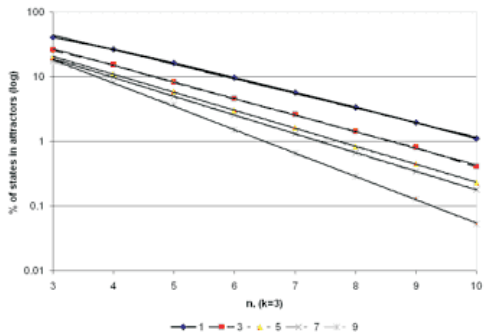,height=4.5cm}}
\caption{Percentage of states in attractors varying maxP (log scale)}
\end{figure}
It is curious to note that for not very small n, the percentage of states in attractors for low maxP increases with
k, but for a high maxP it slightly rises and falls again roughly around $k=3$.

Even for our small statistical samples, percentages of states in attractors fit nicely an exponential curve (Figure
4), as opposed to number of attractors which roughly have linear fit (Figure 2). This suggests that there is less
variance in percentages among networks than in number of attractors.

The tables with the precise data used to generate the graphics  are available through the URL of RBNLab.

\section{Lack of Knowledge about the Context}
Let us now explore the situation where we do not know exactly the context in which a certain DGARBN is. We can
explore a very small RBN and observe its possible pure contexts. We devised a RBN of $n=2$, $k=2$ that makes
explicit the point we want to make. Table 1 shows its transition table.

\begin{center}{\small \em \noindent Table 1. Transition table of a RBN $n=2$, $k=2$.}
\end{center}
\begin{center}
\begin{tabular}{c|c}
Net(t) &   Net(t+1)    \\
\hline
00 &  01\\10 &  10\\01 &  10\\11 &  11\\
\end{tabular}
\end{center}

We consider now pure contexts with maxP=2. The possible dynamics can be seen in Figure 5. Other combinations of
$p$'s and $q$'s give similar attractor basins.
\begin{figure}
\centerline{\psfig{figure=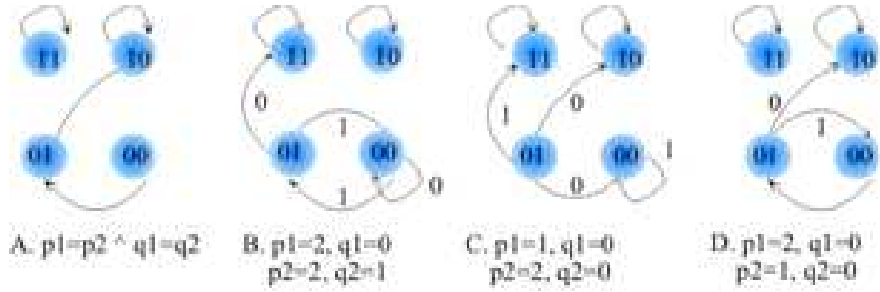, height=3.8cm}}
\caption{Dynamics of a DGARBN $n=2$, $k=2$ with different pure contexts. The arrows
with numbers indicate which transition will take place depending on the modulus of time over two}
\end{figure}
All the pure contexts have at least two attractors: `11' and `10' (only D has an attractor `$01[t mod 2=1] \to 00[t
mod 2=0] \to 01[t mod 2=1]$'). Suppose now the network is initially in some state, e.g. `01', and suppose the
uncertainty on the \emph{mixed} context can be cast in some probabilistic measure; such that each possible
\emph{pure} context is attributed a given weight. When exposing the system to this mixed context, a probabilistic
mechanism  during the exposure of system to context selects one of the pure contexts, but we do not know which one
exactly. Subsequently the network then proceeds in a dynamics according to the selected context. We can say that the
network is in a \emph{potentiality} state, because depending on the exact nature of the mixed context (what the
probability weights on the different pure contexts are), it will go to different attractors with a probability
proportional to the weights of the pure contexts in this specific mixed context. We remark that we know perfectly
the dynamics of the network and the effect of the pure contexts. They are deterministic. There is no lack of
knowledge about the functioning of the network, but there is a lack of knowledge about the pure context, expressed
in the statistical mixture which is the mixed context. As it has been shown, this means that the situation we face
is not a classical mechanical situation, but a quantum-mechanical-like situation \cite{aerts01,aerts02,aerts03,aerts04,aerts05}.

Contextual RBNs of this type could be used to model, for example, how the genes and environment of a cell determine
its behaviour. A stem cell has the \emph{potentiality} of differentiating into almost any cell type. But the
differentiation is not determined only by the genes of the cell, but also by the context (environment) in which it
lays.

\section{Conclusions and Future Work}
We have used deterministic generalized asynchronous random boolean networks to model context change in discrete
dynamical systems. We observed that the context changes dramatically the properties of the systems. An interesting
result was noting that the properties of restricted temporal contexts (small maxP) resemble the dynamics of
synchronous RBNs, while unrestricted ones (large maxP) appear more like non-deterministic asynchronous RBNs.

We introduced two types of context, pure contexts, their interaction with the network is deterministic, and mixed
contexts, more precisely statistical mixtures of pure context, which have a non-deterministic interaction with the
network. The presence of mixed contexts changes completely the dynamics of the network, in the sense that we get
non-deterministic dynamics, generating a probability structure that is non-Kolmogorovian (quantum-like), and
transforming states of the network into potentiality states that collapse with a certain probability to the
attractor states.

For the pure contexts (the deterministic dynamics), it is worth mentioning that in all types of networks, the
percentage of states in an attractor diminishes exponentially (see also \cite{gershenson01}). This means that, \emph{in
theory}, independently of their connectivity ($k$), context, or updating scheme, large networks would always have
Òorder for freeÓ \cite{kauffman01}. In other words, the wide variety of the state space, as $n$ increases the percentage of
states which can ÒtrapÓ the dynamics, \emph{i.e.} states in an attractor, will diminish. The difference in practice
lies in how long will it take to reach this order. In some chaotic networks ($k>2$ for CRBN), this might be infinite
in practice and the order will not be found. Therefore, it would be significant to study of the phase transitions
between order/complexity/chaos in all types of RBNs. Another reason for studying phase transitions, specifically in
contextual RBNs, is to observe if the phase transitions change with the context. In any case, we would expect to
find a complexity region Òat the edge of chaosÓ, but we are interested in finding out wether this also depends on
the context or not.

We also want to study the different regimes under mixed context influence. Remember that the dynamics of our network
becomes non-deterministic in this case, and the probability structure is non-Kolmogorovian and quantum-like. It is
well known that quantum chaos does not exist in the usual manner of classical chaos, because of the fact that in
standard quantum mechanics the dynamics are always linear. The structure that we find here is however not standard
quantum mechanical, it is quantum-like in the sense that the probability model is non-Kolmogorovian, but does not
entail the restriction of linearity \cite{aertsdurt01,aertsvalckenborgh01}. This means that in principle and contrary to
this being the case for standard quantum mechanical systems, we must be able to find chaotic regions. In future work we
want to explore the chaotic, complex, and orderly regimes and the transitions between them under the influence of
non-deterministic contextuality.

The study of the effects of pure contexts in RBNs should also shed light into the debate on the feasability to
correlate the number of human cell types to the attractors of a \emph{contextual} RBN [6, 7, 8, 1].

Finally, the proposed contextual RBNs could be generalized to observe and study the dynamics of discrete
n-dimensional systems interacting with m-dimen\-sional contexts. We could study then not only the state-space of the
system, but also the state-context-space.

\bibliography{CRBN}

\end{document}